\documentclass[aps,preprint]{revtex4}
\input{epsf.tex}
\usepackage{graphicx}
\usepackage{epsf}
\usepackage{wrapfig}
\usepackage{epsfig}
\newcommand{\be}{\begin{eqnarray}}
\newcommand{\ee}{\end{eqnarray}}
\begin{document}

\def\brho{{\mbox{\boldmath$ \rho$}}}

\title{\Large\bf Realistic Calculations of Exclusive  $A(e,e'p)X$ Processes 
off Few-Body Nuclei \footnote{Presented by C. Ciofi degli Atti at the International Workshop on {\bf Probing
Nucleons and Nuclei via the (e,e'p) Reaction}, 14-17 october 2003, Grenoble (France)} }

\author{C. Ciofi degli Atti, 
L.P. Kaptari \footnote{On leave from  Bogoliubov Lab. 
Theor. Phys.,141980, JINR,  Dubna, Russia}}
\address{
    {\it Department  of Physics,  University  of Perugia and INFN, Sezione di Perugia,\\
       Via A. Pascoli, Perugia, I-06100, Italy} }         
\begin{abstract}
  \noindent The exclusive  processes $^2H(e,e'p)n,\, ^3He(e,e'p)^2H$, and  $^3He(e,e'p)(pn)$,
   have been calculated using   realistic few-body wave functions and treating final state interaction  effects
  within a generalized eikonal approach.
\end{abstract}

\maketitle
%
%

\section{Introduction}

\noindent

 One of the main aims of  nowadays hadronic physics is the investigation of the limits  of validity 
 of the so called {\it Standard Model} of
  nuclei, i.e. the description of nuclei through the solution of the non relativistic Schr\"odinger equation containing
  realistic nucleon-nucleon interactions. To this end, 
    exclusive lepton 
  scattering could provide useful information on the nuclear
  wave function,  provided  a reliable  treatment of initial and final states involved in the process
  can be adopted. In the case of few-body systems,  a consistent treatment of initial and final states
   is nowadays 
  possible at low energies by the solution of the Schr\"odinger equation (see e.g. \cite{gloeckle,pisa} and References
   therein quoted).
  However, at high energies,  when the number of partial waves sharply increases and  nucleon excitations  can occur,
   the Schr\"odinger approach  becomes impractical and  other methods
  have to be employed. In this contribution the results of
  calculations of the exclusive process $A(e,e'p)B$, described  within  a generalized  eikonal approach 
  to treat final state interaction (FSI) and  using  realistic 
  few-body wave functions, are reported. 
  
   \section{Basic formulae and results}
    
\vskip 2mm
A. {\it Basic formulas}
\vskip 2mm
   
   In the one-photon-exchange approximation we write the differential cross section of the
   process $A(e,e'p)(A-1)$ in the following form

\be
&&
\frac{d^5\sigma}{dE_{e'}d\Omega_{e'}d{\bf p}_m}
=
K(x,Q^2,{\bf p}_m)\, \sigma_{cc1}^{eN}(Q^2,{\bf p}_m)\, |M_{A,A-1}({\bf p}_m,E_{m})|^2,
\label{eq2}
\ee
 where
 $K(x,Q^2,{\bf p}_m)$ is  a kinematical factor,
 $\sigma_{cc1}^{eN}(Q^2,{\bf p}_m)$  the De Forest CC1 cross section \cite{forest},
 ${\bf p}_m\equiv {\bf q}-{\bf p}'$ the  {\it missing momentum},  i.e.  the Center-of-Mass momentum
of the undetected particles , ${\bf p}'$ the momentum of the detected particle, and  
$E_{m} = \sqrt{P_{A-1}^2}+M_N-M_{A} = q_0-T_{p'} - T_{A-1}$  the {\it missing energy}.

\begin{figure} 
\hspace*{-2mm}                     
\epsfig{file=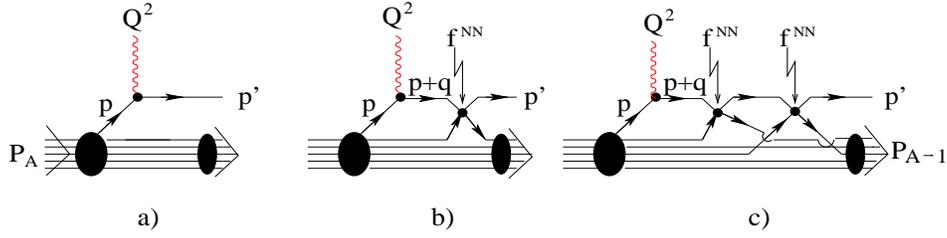,width=12.5cm,height=4.0cm}
\caption{ The Feynman diagrams for the process $A(e,e'p)(A-1)$: the  Plane Wave Impulse Approximation (PWIA) $a)$, and the
 single $b)$  and
double $c)$ rescattering in the final state. $f^{NN}$ denotes the elastic nucleon-nucleon (NN) scattering amplitude.}
   \label{fig1}
 \end{figure}

 In our approach, the nuclear transition matrix element $M_{A,A-1}({\bf p}_m,E_{m})$ in eq. (\ref{eq2}) 
 is computed  by evaluating the corresponding Feynman diagrams of
  Fig. \ref{fig1}, which describe the 
 interaction of the incident electron with one nucleon of the target 
followed by its elastic  rescattering with the   nucleons of the $(A-1)$  nucleus.

\vskip 2mm
B. {\it The process  $^2H(e,e'p)n$.}
\vskip 2mm
  
 In PWIA (Fig 1a)) the transition matrix element simply becomes  the deuteron momentum distribution, i.e.
  
 \be
 | M_{A,A-1}|^2 \to n_{D}( {\bf p}_m) =
\frac13\frac{1}{(2\pi)^3} \sum\limits_{{\cal
M}_D} \left | \int\, d  {\bf r} \Psi_{{1,\cal
M}_D}( {\bf r})  \chi_f\,\exp (-i{\bf p}_m {\bf r}) \right |^2
\label{deut}
\ee
   
\begin{figure}[t]                     
\begin{minipage}{6cm}
\epsfig{file=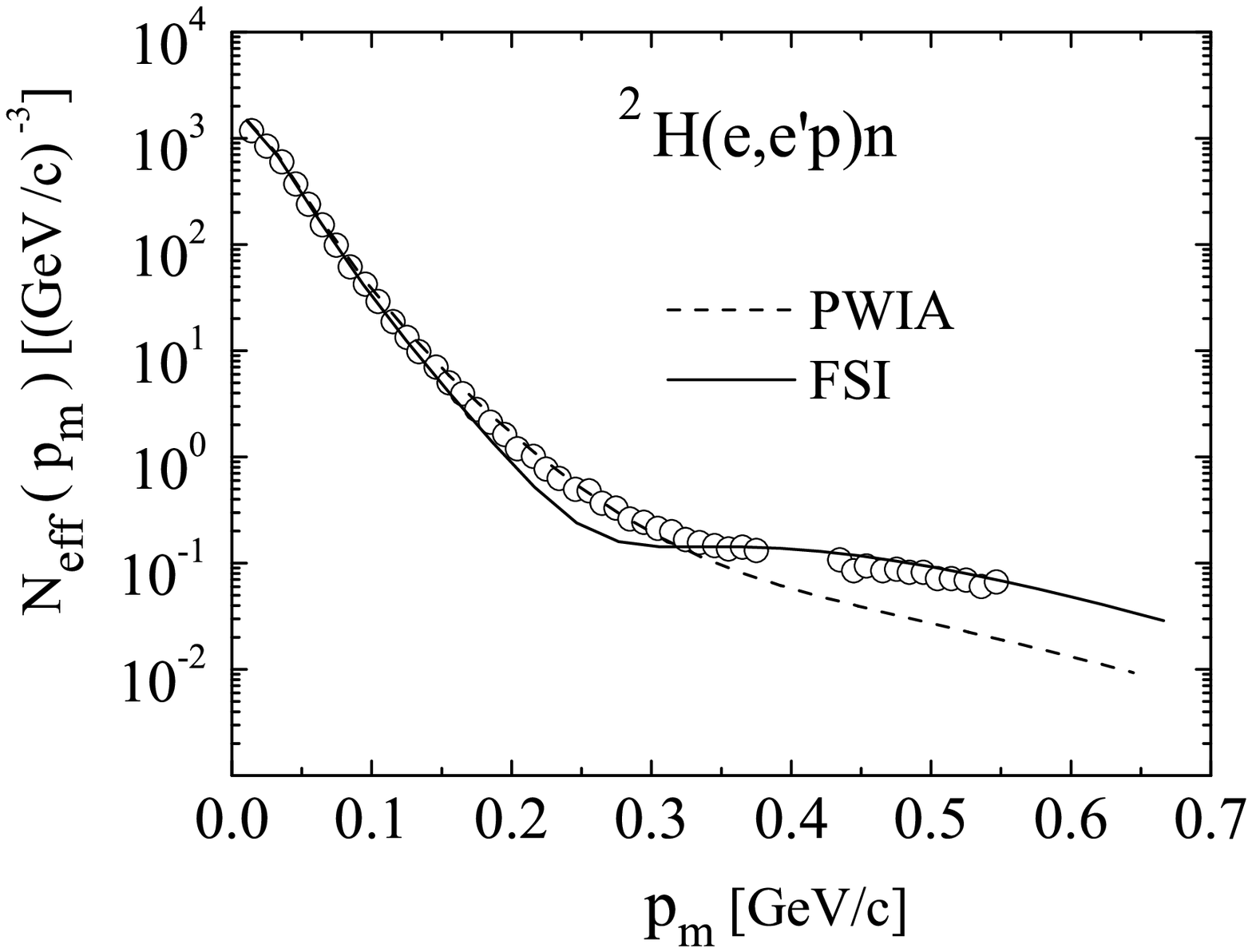,width=5.60cm,height=4.80cm}
\end{minipage}
\begin{minipage}{6cm}
\vskip 1mm
\epsfig{file=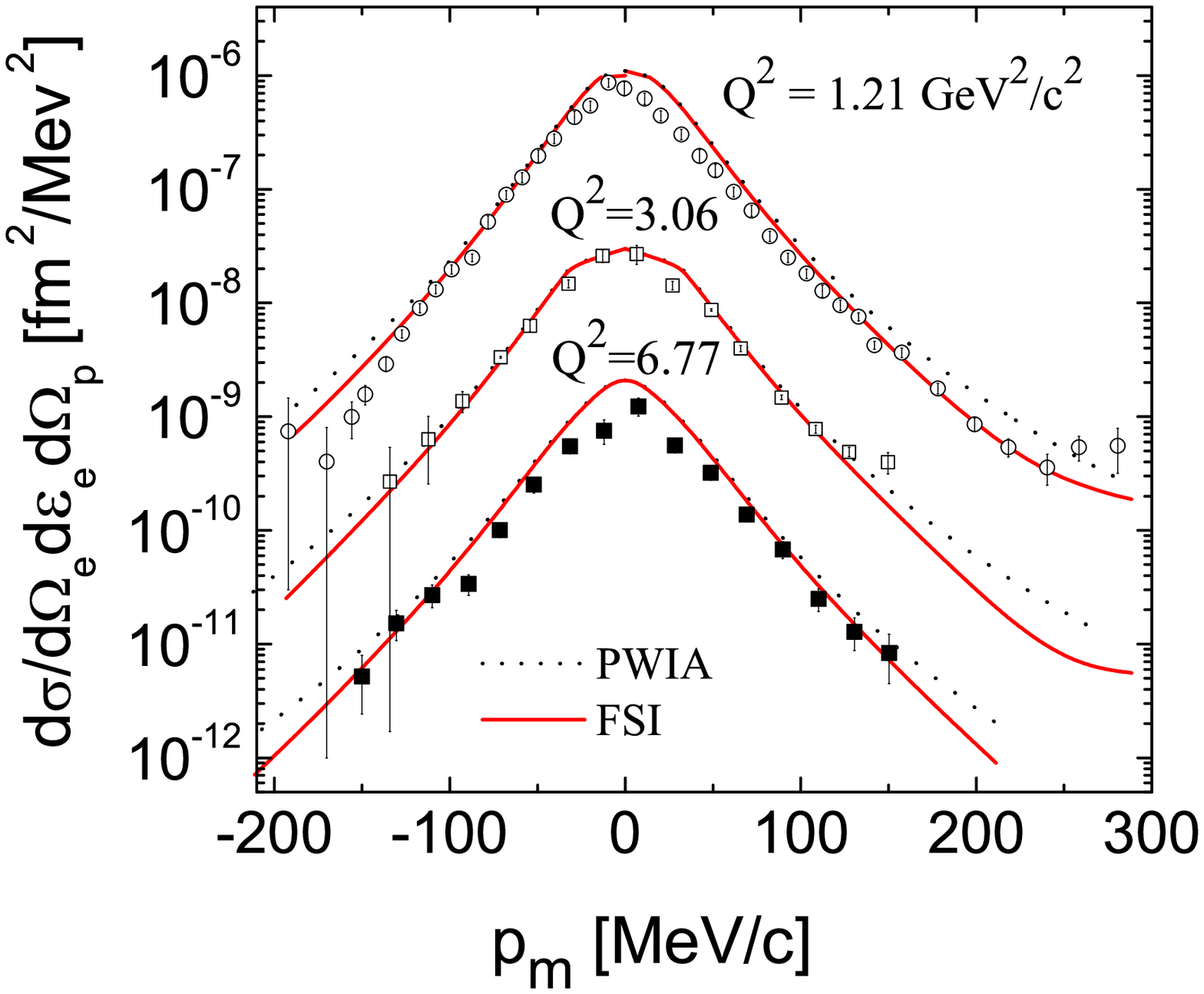,width=5.60cm,height=4.750cm}
\end{minipage}
\caption{ Comparison of theoretical calculations with the experimental data from JLAB \protect\cite{ulmer} (left) 
and SLAC \protect\cite{bulten}
(right). The negative values of 
\protect $p_{m}$ correspond to protons  detected at $\phi=0$.}
   \label{fig2}
 \end{figure}

When the relative energy of the  $np$-pair is large,
  the exact two-nucleon continuum wave function can be approximated by its  eikonal  form,
obtaining \cite{w}

\begin{equation}
n_{D} \to N_{eff}({\bf p}_m) =
\frac13\frac{1}{(2\pi)^3} \sum\limits_{{\cal
M}_D} \left | \int\, d  {\bf r} \Psi_{{1,\cal
M}_D}( {\bf r}) S( {\bf r}) \chi_f\,\exp (-i
{\bf p}_m {\bf r}) \right |^2, \label{ddistr}
\end{equation}
where $
{\cal S}({\bf r}) =
 \left[ 1-\theta(z)\Gamma({\bf b})\right ]
$ and 
 $z$ and ${\bf b}$ are  the longitudinal and transverse co-ordinates  with respect to
the direction of the struck nucleon.
 In Fig. \ref{fig2} the results of our calculations are compared with
 the experimental data (here, and in what follows, we used
$
\Gamma({\bf b}) =\displaystyle{\sigma_{NN}^{tot}[(1-i\alpha)}/(4\pi b_0^2)]
exp(-{\bf b}^2/2b_0^2)$, 
with the values of the total cross section $\sigma_{NN}$, the ratio
$\alpha = Re f^{NN}(0)/Im f^{NN}(0)$ and the slope parameter $b_0$ from \cite{experi}).
 The right panel in Fig. \ref{fig2}
 illustrates the $Q^2$ dependence of the cross section 
 at two different values of  the azimuthal angle $\phi$ between
 the scattering and reaction planes, 
 namely $\phi=0$ (negative values of $p_m \equiv|{\bf p}_{m}|$)
 and $\phi=\pi$ (positive values of $ p_m$).      
It can be  seen that FSI effects lead always  to a better agreement with the experimental data.

\vskip 2mm
C. {\it The processes  $^3H(e,e'p)D$ and $^3H(e,e'p)pn$}
\vskip 2mm
 
 For a  $^3He$ target, the final state can be   either the deuteron  or the continuum two-body state.
 We have considered the following three cases:\\
 \noindent  
  1. {\it the PWA approximation}: all particles in the final states are described by 
  plane waves, which means that  the transition matrix element is nothing but the three-body ground state wave function 
  in momentum space;
  
 \noindent
  2. {\it the PWIA}: in this picture (Fig.\ref{fig1}a)) the struck proton is always 
  described by a plane wave and the FSI is only taken into account in the $(np)$ pair
  of the  three-body channel process
  $^3He(e,e'p)(np)$; 
 in our calculations both the two- and three-body  wave functions
 correspond to the $AV18$ interaction \cite{av18}, with the three-body wave function from \cite{pisa}.\\ 
\noindent
3. {\it the full FSI}: the  
$(np)$ system (ground or continuum states) is still described by  the exact
solution of the Schr\"odinger equation, whereas the  interaction of the 
 struck nucleon with the pair is treated  by evaluating the Feynman diagrams of Figs. 1 b) and 1c).
For the three-body channel, one obtains
    \be 
     \label{Hepnn}
      && |M_{A,A-1}|^2 \equiv P_D({\bf p}_m, E_m) = \int d {\bf k}\\\nonumber 
 &&\left | \int\, d {\bf r} d {\brho} \phantom{\frac12}     \!\!
   \Psi_{^3He} ({\bf r}, {\brho}) {\cal S}^{FSI}(\brho,{\bf r})  
   \exp (i{\bf p}_m \brho)  \ \phi_{12}^{{\bf k}}({\bf r})
     \right |^2 \delta\left( E_{m}-E_{3}-\frac{{\bf k}^2}{M_N}\right)
     \ee 
where 
$
{\cal S}^{FSI}({\bf r}_1,{\bf r}_2,{\bf r}_3) =
\prod\limits_{i=1}^{2}\ \left[ 1-\theta(z_i-z_3)\, \rm e^{i{\mbox{$\Delta_0(z_i-z_3) $}}}\Gamma({\bf b}_i-{\bf b}_3)\right ]
$
and $\Delta_0\sim (q_0/|{\bf q}|) E_{m}$ is a  factor which appears when the frozen approximation underlying 
the Glauber approach is released and the recoil momentum of the third nucleon, appearing when the struck
nucleon rescatters on the second one, is taken into account\,\,\cite{glauber,misha}. The effects from the factor
$\Delta_0$ increase with the removal energy,  but in most cases considered  they do not appreciably distort the Glauber result
(this point is still under investigation \cite{misha}).
The transition matrix element 
for the two-body channel has the same form, with the continuum two-body wave function replaced by the deuteron
wave function, and the argument of  the energy-conserving  $\delta$-function
 properly modified. In eq. \ref{Hepnn},  $P_D({\bf p}_m, E_m)$ represents the  {\it distorted}
  Spectral Function, which, 
  when $\Gamma = \Delta_0 = 0$, reduces to the usual one 
 $P(|{\bf p}_m|, E_m)$  \cite{claleo}.

 \begin{wrapfigure}[17]{l}[0pt]{6.cm}           
\epsfig{file=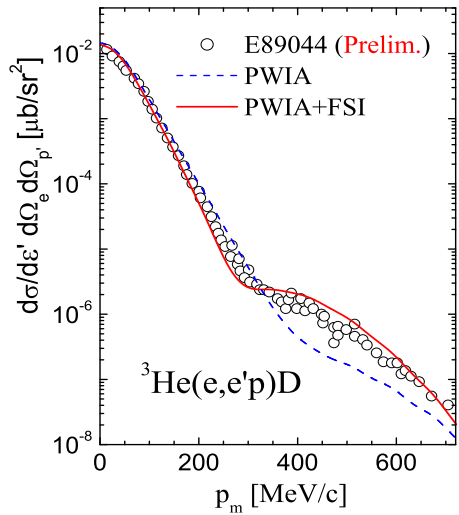,width=6.0cm,height=6.9cm}
\vspace*{-0.5cm}
 \caption{Comparison of our  calculations (AV18 interaction) of the
  two-body channel process  with preliminary experimental data from \protect\cite{jlab}.
  Three-body wave function from~\protect\cite{pisa}.}
 \label{fig3}
  \end{wrapfigure}
\begin{figure}[ht]           
\begin{center}
 \epsfig{file=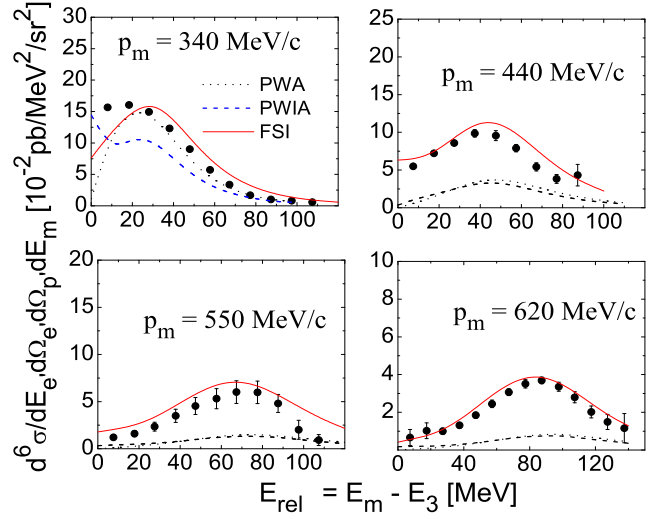,width=8.7cm,height=7cm}
 \end{center}
\caption{Comparison of our  calculations (AV18 interaction) of the process   $^3He(e,e'p)np$ 
with preliminary experimental results from 
 \protect\cite{jlab}. Three-body wave function from
\protect\cite{pisa}.}
\label{fig4}
\end{figure}

   The results of our calculations are shown in Figs. \ref{fig3} and  \ref{fig4}.
Both sets of data refer to the {\it  perpendicular kinematics}, when
the final  proton is detected  almost perpendicularly to  ${\bf p}_m$; it can be seen that  in the two-body channel process,  the inclusion of 
FSI effects, appreciably improves the agreement with the experimental data. As for the three-body channel,
one sees that  at sufficiently high values of $E_m$ and $p_m$, the $PWA$ and $PWIA$ predictions practically
coincide, in agreement with the behaviour of the Spectral Function which, as shown in Fig. 5 (see  \cite{claleo}),
exhibits bumps at $E_m \simeq {\bf p}_m^2/(4m)$ originating from two-nucleon correlations. Thus, if the PWIA
were valid, the $^3He(e,e'p)(np)$ cross section at $p_m \geq 440 MeV/c$ and  $E_m \geq 10 MeV$ would be directly related
 to the
three-body wave function. Unfortunately, one sees  that in the perpendicular kinematics of~\cite{jlab},  the FSI between 
the struck proton and the $(np)$ pair almost entirely exhausts the cross section. However, as shown in Fig. 5, this does not
seem to be the case for the experimental data of \cite{saclay}, where  the struck nucleon is detected
almost along the direction of \, ${\bf p}_m$.
\section{Conclusions}
We have developed a realistic approach aimed at a consistent treatment of  initial state correlations and 
FSI effects in exclusive
$A(e,e'p)X$ processes; the approach is  based upon the use of realistic three-body wave functions
 corresponding to the AV18 interaction and a generalized
eikonal approach, where the Glauber frozen approximation is released. Our results show that by a proper choice of the
 kinematics,  FSI effects might   appreciably be 
reduced, as also occurs  in the $^3He(e,e'2p)n$ process (see in particular Fig.10 of\,\, \cite{claleo});
 thus it appears that  by
quasi elastic  exclusive processes, the details of the ground-state few-body wave function can eventually be investigated.

 \begin{figure}[ht]                    
\begin{minipage}{7cm}
   \includegraphics[width=60mm,height=6cm]{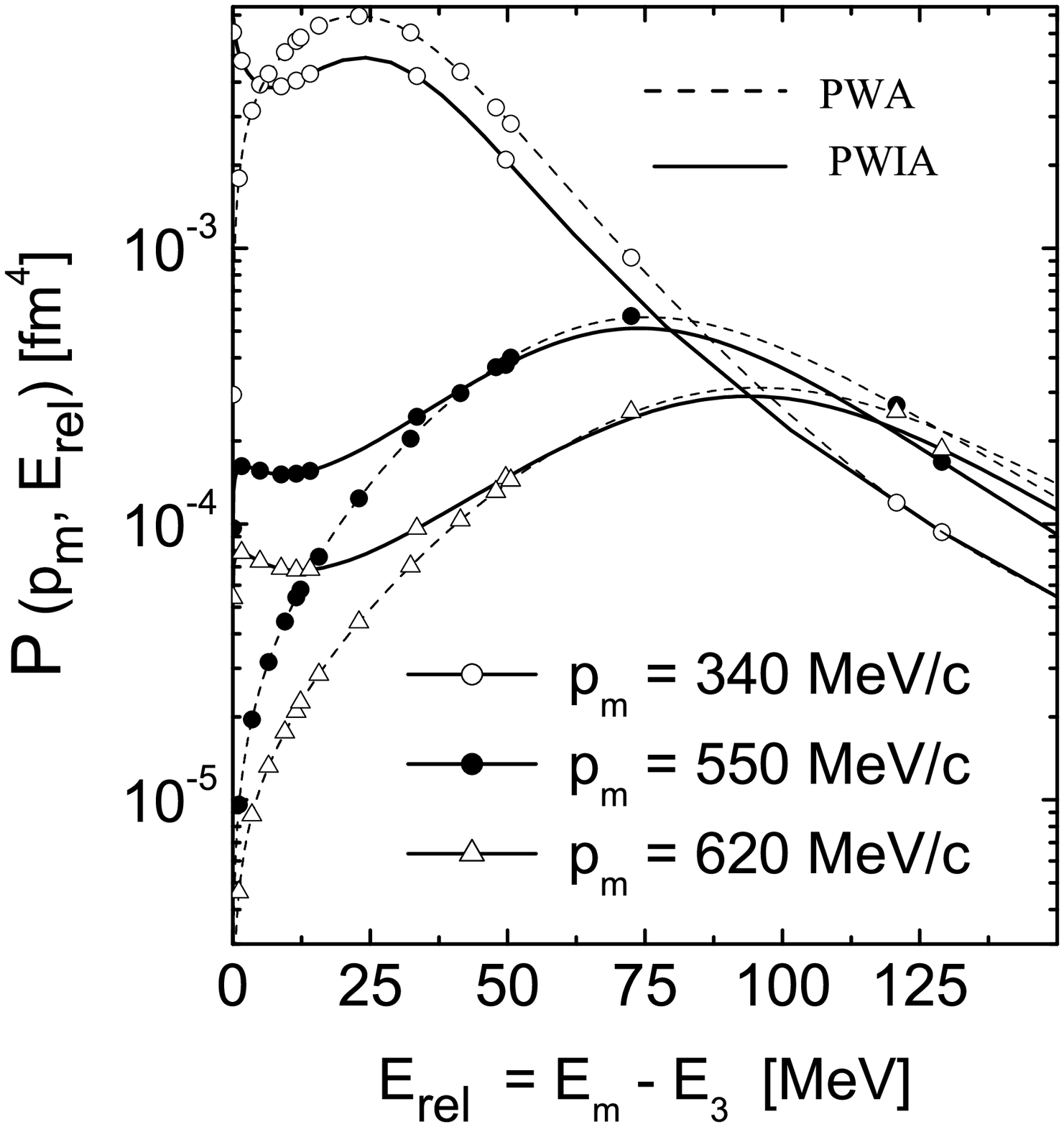}
   \end{minipage}
\begin{minipage}{7cm}
\includegraphics[width=50mm,height=7.5cm]{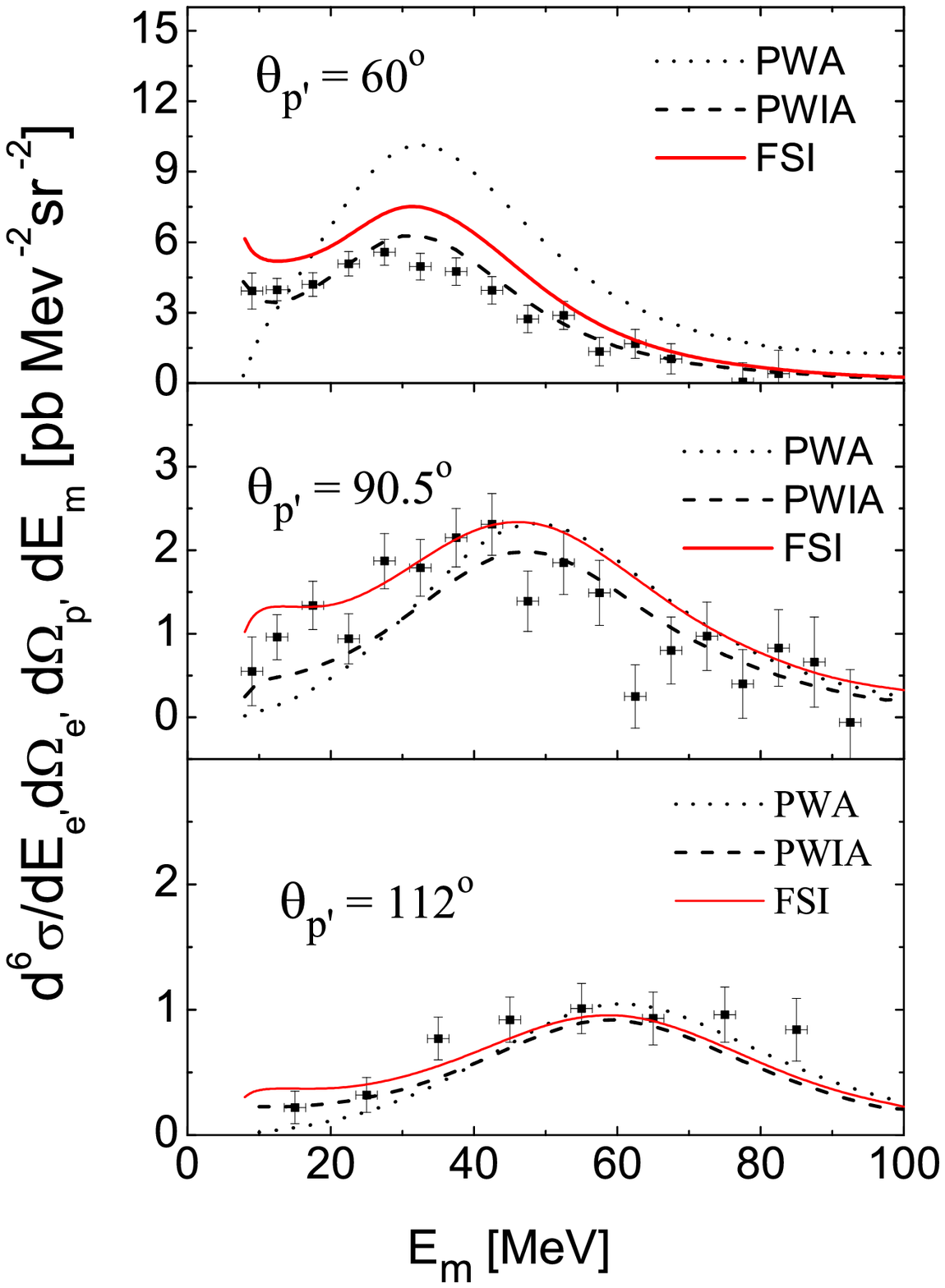}
   \end{minipage}
\caption{ {\it Left panel}: the proton Spectral Function (AV18 
interaction) of $^3He$ \protect\cite{claleo}. {\it Right panel}: 
comparison of our theoretical calculations (AV18 interaction) of the process  $^3He(e,e'p)np$ 
with the  results from  \protect\cite{saclay}. Three-body wave function from
\protect\cite{pisa}.} 
\label{fig5} 
\end{figure}%

\newpage
\section{ Acknowledgments}
We are grateful to A. Kievsky for providing the Pisa group three-body wave functions and to the staff of the Jlab E-89-044
experiment for  information on their preliminary experimental data.

%

\end{document}